\documentclass[sigplan,10pt,nonacm]{acmart}         
\makeatletter                   
\def\mdseries@tt{m}             
\makeatother        

\makeatletter
\renewcommand\@formatdoi[1]{\ignorespaces}
\makeatother

\usepackage[export]{adjustbox}
\usepackage{listings}
    
\usepackage[newfloat=true,frozencache=true]{minted} 
\usepackage{color}                                                                   
\usepackage{xcolor}                                                                  
\usepackage{changepage} 
                                                                                     
\usepackage{url}                                                                     
\hypersetup{hidelinks=false,colorlinks=true,linkcolor=blue,urlcolor=blue,citecolor=blue,anchorcolor=blue} 
\usepackage{breakurl}                                                                

\usepackage{tabularx}
\usepackage{booktabs} 

\usepackage{cleveref}
\crefformat{section}{\S#2#1#3}
\crefformat{subsection}{\S#2#1#3}
\crefformat{subsubsection}{\S#2#1#3}
\crefrangeformat{section}{\S\S#3#1#4 to~#5#2#6}
\crefmultiformat{section}{\S\S#2#1#3}{ and~#2#1#3}{, #2#1#3}{ and~#2#1#3}

\usepackage{microtype}
                                                 
\usepackage[utf8]{inputenc}                                                          

\usepackage[iso,american,cleanlook]{isodate}    

\usepackage{rviewport}                                                               

\usepackage{tikz} 
\usepackage[inline]{enumitem} 

\newcommand\Invisible[1]{                                                            
  \marginpar{\color{white}{\fontsize{.5}{.5}\selectfont #1 }}                        
}

\newcommand{\Exclude}[1]{}

\newcommand\ParaBold[1]{\vspace{0.5 \baselineskip} \noindent \textbf{#1} \noindent}

\definecolor{Gray95}{gray}{0.95}
                                                                                     
\newcommand{\AtFoot}[1]{\let\thefootnote\relax\footnotetext{{#1}}}

\startPage{1}

\acmConference[]{}{}{}
\acmYear{}
\acmISBN{} 
\acmDOI{} 
\startPage{1}

\usepackage{booktabs}   
\usepackage{subcaption} 

\newcommand{\orcidicon}[1]{\href{https://orcid.org/#1}{\includegraphics[scale=0.06]{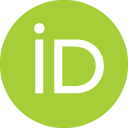}}}

\begin{document}

\title[]{Semaphores Augmented with a Waiting Array} 


\author{Dave Dice \orcidicon{0000-0001-9164-7747}}
\orcid{0000-0001-9164-7747}             
\affiliation{
  \institution{Oracle Labs}             
\country{USA}                    
}
\email{first.last@oracle.com}            

\author{Alex Kogan \orcidicon{0000-0002-4419-4340}} 
\orcid{0000-0002-4419-4340} 
\affiliation{
  \institution{Oracle Labs}             
\country{USA}                   
}
\email{first.last@oracle.com}          


\begin{abstract}

Semaphores are a widely used and foundational synchronization and coordination construct used for
shared memory multithreaded programming.  
They are a keystone concept, in the sense that most other synchronization
constructs can be implemented in terms of semaphores, although the converse does not generally hold.  
Semaphores and the quality of their implementation are of consequence as they remain heavily used in the
Linux kernel and are also available for application programming via the \texttt{pthreads} programming
interface.  

We first show that semaphores can be implemented by borrowing ideas from the classic
\emph{ticket lock} algorithm.  The resulting \emph{ticket-semaphore} algorithm is simple and compact 
(space efficient) but does not scale well because of the detrimental impact of global spinning.  
We then transform \emph{ticket-semaphore} into the {TWA-semaphore} by the applying techniques
derived from the \emph{TWA - Ticket Locks Augmented with a Waiting Array} algorithm, yielding
a scalable semaphore that remains compact and has extremely low latency.

\end{abstract}

\ccsdesc[300]{Software and its engineering~Multithreading}
\ccsdesc[300]{Software and its engineering~Mutual exclusion}
\ccsdesc[300]{Software and its engineering~Concurrency control}
\ccsdesc[300]{Software and its engineering~Process synchronization}


\keywords{Semaphores, Locks, Mutexes, Mutual Exclusion, Synchronization, Concurrency Control}  

\maketitle

\thispagestyle{fancy}
\renewcommand{\footruleskip}{.7cm}
\fancyfoot[CF]{\vspace{.7cm} \thepage \\ \today \hspace{1mm} \textbullet \hspace{1mm} Copyright Oracle and or its affiliates} 

\Invisible{Oracle patent disclosure accession number : IDF-138780}

\section{Introduction}

Semaphores are perhaps the oldest known synchronization construct, dating back to at least 1962 or 1963 
\cite{Semaphore-Dijkstra,History}.  Semaphore constitute a key building block -- many other synchronization constructs, 
such as locks (mutexes) or condition variables can and are be trivially implemented in terms of semaphores.  
Semaphores are also commonly used to build higher level primitives, such as thread pools, bounded buffers, 
producer-consumer patterns, reader-writer locks, messages queues, and so on \cite{BookOfSemaphores}.  

Conceptually, a semphore could be implemented as a simple atomic counter.  The \texttt{take} operator
waits for the counter to become positive, and then decrements the counter and returns. 
The \texttt{post} operator increments the counter  
\footnote{The naming conventions for semaphore operators vary considerably.  
Take-Post are sometime called P-V, acquire-release, consume-produce, wait-signal, down-up, get-put, procure-liberate, 
procure-vacate, proberen-verhogen, etc.}. 

In practice, such an implementation, while useful
for explication, is not generally considered viable -- it fails common quality of implementation (QoI) critera. 
Such an implementation, for instance, does not guarantee first-come-first-served admission order,
which is a desirable property.

Semaphores are widely available. The POSIX \texttt{pthreads} application programming interface, C++20,  
and Java's \texttt{java.\allowbreak{}util.\allowbreak{}Concurrent} expose semaphore implementations
for application use. The linux kernel also makes heavy internal use of semaphores.  

\ParaBold{Ticket Locks} We briefly shift our attention to \emph{ticket locks} \cite{focs79,cacm79-reed,tocs91-MellorCrummey}, 
which are a mutual exclusion primitive, and
show how ticket locks can be transformed into semaphores.  
A ticket lock consists of \texttt{ticket} and \texttt{grant} fields.  Threads arriving to acquire
the lock atomically fetch-and-add to advance the \texttt{ticket} value and then wait for the 
\texttt{grant} field to become equal to the value returned by the fetch-and-add primitive.  
The unlock operator simply increments the \texttt{grant}.  Incrementing \texttt{grant} does not 
require an atomic read-modify-write operation.  

\ParaBold{Ticket-Semaphore} We can transform a ticket lock in a semaphore, yielding
the \emph{Ticket-Semaphore} algorithm, by making the following changes to the ticket lock algorithm.
First, we require the use of an atomic increment in update the \texttt{grant} field, as multiple
threads might release the semaphore concurrently. (With a mutex, in contrast, only a single thread,
the owner, can release the lock at any given time).  Next, we use \emph{magntitude}-based comparisons
in the \texttt{take} operator instead of simple equality comparisons.  If the unique ticket value,
obtained from the atomic fetch-and-add operator exceeds the value of \texttt{grant} field, then
the thread can proceed, otherwise it needs to wait until that condition is satisfied. 
And finally, as we are using 
magnitude-based comparisons, arithmetic roll-over of the \texttt{ticket} and \texttt{grant} becomes a concern.
To address that issue we simply ensure that \texttt{ticket} and \texttt{grant} are 64-bit unsigned integers.  
Assuming a processor could increment a value at most once per nanosecond, these fields would not overflow
and wrap around in less than 200 years, so arithmetic overflow is not a practical concern.  
Like ticket locks, \emph{Ticket-Semaphore} provides strict first-come-first-served admission order,
assuming the underlying atomic fetch-and-add primitive is wait-free.  
Listing \ref{Listing:Ticket-Semaphore} illustrates a canonical embodiment of the \emph{Ticket-Semaphore} algorithm.

\Invisible{allay concerns about overflow}
\Invisible{Monotonically increasing; no retrograde movement} 
\Invisible{Anticipatory; anticipate; warmup; staged; staging; pipelined; FSM; phased; modal; 
tee up; standby; signal; inform; poke; nudge; alert; cue; prime; prod; } 
\Invisible {Threshold; LongTermThreshold; NHot; Epsilon; Delta} 
\Invisible {LongTermThreshold; Anticipatory; forewarn; Early-warning; tip-off; clue; 
preemptive warning; Prepare; predictive; anticpatory warmup; make-ready; } 

While this semaphore is simple and compact, and offers extremely low latency absent contention or at low contention,
it suffers, like ticket locks, from excessive coherence traffic generate by \emph{global spinning} where multiple threads 
busy-wait on the \texttt{grant} field, and thus performance fades as we increase the number of participating threads. 

\ParaBold{TWA} TWA (Ticket Locks Augmented with a Waiting Array) \cite{TWA-EuroPar-2019, TWA-arxiv} provides a way to
make ticket locks scale, while still remaining compact and retaining the desirable low handover latency
found in ticket locks.  Under TWA, threads arriving at the \texttt{lock} operator atomically fetch-and-add
$1$ to the \texttt{ticket} field to obtain their private ticket value.  The thread then compares that ticket value
in hand to the \texttt{grant} field.  When \texttt{grant} becomes equal to the ticket value, the thread has obtained
ownership of the lock and can enter the critical section, otherwise it waits.  If the numeric difference
between \texttt{grant} and the ticket value is small, under some threshold, then we allow the thread to use
global spinning on \texttt{grant} but otherwise if the difference exceeds the threshold then we force 
the thread to switch to \emph{semi-local} longer-term waiting as follows.
The thread hashes the address of the lock and the ticket value to form an index into a new \texttt{waiting array}
that is used for longer term waiting.  The \texttt{unlock} operator increments \texttt{grant} as usual,
but then also hashes the \texttt{grant} value to compute an index into the waiting array,  and
updates the waiting array element to notify specific long term waiters that they need to recheck the distance between their
ticket value and the \texttt{grant} field, shifting from long-term waiting to short-term on \texttt{grant} as 
appropriate.  Absent hash collisions, this approach notifies just the single corresponding waiter. 

Crucially, this approach greatly reduces global spinning, to at most one thread per lock at a given time,
and diffuses waiting over the waiting array via the hash operation.  
The hash operator is intentionally designed to be ticket-aware.  The \texttt{waiting array}
is shared by all threads in the process and is of fixed size.  Because of hash collisions, multiple threads might wait on the same
\texttt{waiting array} element, thus we have statistical \emph{semi-local} waiting instead of strict local waiting.
Lock handover is efficient as the successor thread is waiting directly on the \texttt{grant} field.
When the \texttt{unlock} operator increments \texttt{grant}, the successor can notice immediately and enter
the critical section.  While the successor is entering the critical section, the \texttt{unlock} operator, in parallel,
signals the successor's successor (if any) via the long-term \texttt{waiting array} to shift from long-term waiting on
the \texttt{waiting array} to short-term waiting on \texttt{grant}.  Cueing and staging of that next thread overlaps with 
entry and execution of the critical section, allowing for useful parallelism. 

The tunable threshold parameter is named \texttt{LongTermThreshold} in our source code listings, 
and we use $1$ as a default value, but note that the optimal value is platform-specific and related
to the overheads imposed by global spinning, which is driven by coherent cache communication costs.  
Empirically, we have found that $1$ works well under a wide variety of platforms.  

\Invisible{Pipeline; pipelined; Cue; staging; phased; overlap; modal}  

\ParaBold{TWA-Semaphore} Just as we transformed a simple non-scalable ticket lock into the scalable TWA variant,
we can, by analogy, apply those same lessons to transform \emph{Ticket-Semaphore} into a 
scalable form, which we call \emph{TWA-Semaphore}.  
The resultant \emph{TWA-Semaphore} is fair, with a first-come-first-served (FCFS), or,
more precisely, first-come-first-enabled \cite{FIFE} admission order.  
We show a working example in Listing \ref{Listing:TWA-Semaphore}.  We note that this
transformation is readily applicable to other synchronization constructs, such as \emph{EventCount and Sequencers}
\cite{cacm79-reed}.


\section{Waiting Strategies}  

A critical design choice for both locks and semaphores is exactly \emph{how} threads wait in the \texttt{take} 
primitive.  A detailed survey of waiting strategies appears in \cite{arxiv-Malthusian}.  We describe a set 
selected subset of approaches suitable for our semaphore designs.  

\ParaBold{Spinning} Also called \emph{busy-waiting}, threads simply loop, fetching (polling) and checking variables 
until the condition of interest is satisfied.  Spinning loops are usually decorated with a \texttt{PAUSE} or \texttt{YIELD}
instruction which informs that processor that a busy-waiting is active.  Such instructions may act to cede resources
to other processors, reduce thermal or energy consumption, and inform hypervisors (virtual machine monitors) that busy-waiting
is in play to enable gang scheduling, if necessary.  \texttt{PAUSE} is considered an advisory ``hint'' to the system and
has no specific semantics.

Loops may also include calls to operating system services such as \texttt{sched\_yield}, which hints to the scheduler
to transiently surrender the processor to other potentially runnable threads. The actual semantics 
of \texttt{sched\_yield} are intentionally defined to be advisory and vague, providing considerable latitude
to the implementation.  For instance lets say or system has 2 CPUs.  Thread $T1$ is running on CPU-0 and thread $T2$ is 
running on CPU-1.  Thread $T3$ is in ready state on CPU-0's local ready queue.  If $T2$ calls \texttt{sched\_yield}
there is no guarantee that $T3$ will be scheduled onto CPU-1.  Often, the call to \texttt{sched\_yield} will return
immediately if there are no ready threads on the caller's local ready queue.  As such, \texttt{sched\_yield} is not 
particularly helpful for spin loops.  

While waiting, the spinning thread occupies a CPU
that might be otherwise used by other ready threads.  (And in fact those other ready threads might be the ones
destined to satisfy the condition of interest).  As such, except for certain kernel environments, long-term unbounded spinning is
frowned up and is considered an \emph{anti-pattern}.  
In particular, if we have more runnable threads than processors, then involuntary preemption,
as provided by the scheduler, can be deeply problematic with respect to performance. 
For instance, lets say our system has 4 processors.  
Threads $T1$, $T2$, $T3$ and $T4$ have been dispatched onto the processors are busy-waiting for a lock held by
$T5$ where $T5$ is runnable but has been preempted.  To make forward progress we need to $T5$ to be scheduled via 
preemption of one of the other threads.  Preemption, however, tends to operate in longer millisecond time frames, so
throughput over the lock can be greatly impaired by using a pure spinning waiting strategy.  

In the context of \emph{TWA-Semaphore}, spinning could be used both short-term spinning on the \texttt{grant} field
or longer-term waiting via the waiting array.  

\ParaBold{MONITOR-MWAIT} on the x86 architecture, and similar facilities, such as \texttt{WFE} or \texttt{WFET} on ARM processors, offer 
a more ``polite'' mode of waiting, potentially consuming less resources during polling, but still cause 
the waiting thread to occupy a processor. 

\ParaBold{Linux Kernel Futex} The \emph{futex} \cite{futex}  mechanism allows threads to designate an address of interest and then
block in the kernel, surrendering the processor (allowing other threads to immediately use the processor), 
until some other thread performs a corresponding notification system call on that same address.  
Microsoft Windows exposes a similar \texttt{WaitOnAddress} mechanism.  
The underlying \texttt{futex} implementation hashes the specified virtual address into an index into a 
kernel hashtable, where each bucket contains a spinlock and a pointer to a linked list of threads waiting on that bucket.  
The spin lock is used to protect that specific bucket.  

In the context of \emph{TWA} or \emph{TWA-Semaphore}, we could utilize the \texttt{futex} services to wait on the 
buckets in our waiting array.   This approach confers an additional scalability advantage.  As the \texttt{futex}
operator hashes address into kernel hashtable buckets, but by using multiple address in the waiting array, we in turn
disperse \texttt{futex} operations over the entire futex hashtable, avoiding hot spots and reducing contention
on individual per-bucket futex hash table spinlocks.  

Threads deeper on the logical queue than \texttt{LongTermThreshold} wait via \texttt{futex} while those
closer to the front spin.  If we desire that all threads wait by \texttt{futex} and need to eliminate all
spinning, then we simply set \texttt{LongTermThreshold} to 0.  We note too that hybrid \emph{spin-then-park}
\cite{arxiv-Malthusian} waiting strategies could also be employed, where threads would a attempt a brief bounded local spinning 
phase followed, if necessary, by \texttt{park}.  The goal is to avoid voluntary context switching costs
imposed by \texttt{park-unpark} if the waiting period is short.  

\ParaBold{Waiting Chains} Some low-level waiting primitives, such as \texttt{park-unpark} \cite{LEA2005293}, 
allow a thread to \texttt{park}, which deschedules the calling thread until a corresponding \texttt{unpark}.  
Again, this allows the CPU of thread calling \texttt{park} to be reallocated and dispatched onto immediately,
allowing other ready threads to start running.
If the \texttt{unpark} were to execute before the corresponding \texttt{park}, the threading system maintains 
a per-thread flag set accordingly, and the subsequent \texttt{park} operation clears
the flag and returns immediately. \texttt{Park-unpark} can thus be considered to provide a bounded binary per-thread
semaphore.  \texttt{park-unpark} is a \emph{1:1} point-to-point model, in that the \texttt{unpark} operator must know and pass
the identity of the specific thread to be unparked, and \texttt{unpark} wakes just that one thread.    

We note that \texttt{park-unpark} is \emph{identity-based} while the \texttt{futex} operator is address-based.  
Using waiting chains, we can easily implement address-based waiting via identity based park-unpark primitives.  

To use \texttt{park-unpark} services with \emph{TWA} and \emph{TWA-Semaphore} we propose the following algorithm,
described in detail in Listing \ref{Listing:TWA-Semaphore-Chains}.  In this case, the waiting array elements are
pointers to linked lists (stacks) of threads waiting at that bucket.  Waiting threads use an atomic
\texttt{exchange} operator to push \texttt{Waiting Element} addresses onto the stack.  The waiting elements
can be allocating on the calling thread's stack and require no special memory management considerations.  
The elements contain a flag (\texttt{Gate}) and, if using \texttt{park-unpark}, the identity of the 
associated thread.  To notify long-term waiters at a particular index, we use an atomic \texttt{exchange}
to detach the entire chain (stack) and then \texttt{unpark} all the threads on the chain, allowing them
the opportunity to reevaluate their corresponding \texttt{grant} values to see if the wait condition
has been satisfied.  Critically, management of these chains is itself completely lock-free.   
We have implemented a concurrent \emph{pop-stack} \cite{pop-stack} to maintain lists of waiting threads. 
For brevity, in Listing \ref{Listing:TWA-Semaphore-Chains} we have elided the \texttt{LongTermThreshold} logic.

We observe that the optimized waiting techniques described in \cite{DelegatedConditionVariables}, where
the condition of interest is (re)evaluated by the notifier instead of the notifiee, are applicable 
in this context, and would allow further optimizations to the waiting strategy.  

Using waiting chains may still be useful even if \texttt{park-unpark} is not available.  In that case \texttt{park} calls
could be replaced by \texttt{PAUSE} and \texttt{unpark} becomes an empty ``no-op'' operation, in which 
case waiting devolves to unbounded local spinning.  

We also note that waiting chains offer another avenue for optimization.  When thread in \texttt{post} performs 
notification, we can easily propagate the address of the semaphore and the posted \texttt{grant} value along the
notification chain, from waiter to waiter.  When a waiter wakes, it can then check its own condition against
that passed information, which in turn can often allow the waiter to recognize that it has been admitted without
any need to fetch and check against the \texttt{grant} field, further reducing coherence traffic on the central
semaphore shared variables.

\Invisible{ ORA200224-US-NP : Efficient Condition Variables via Delegated Condition Evaluation
https://patents.google.com/patent/US20210311773A1}  

\Invisible{accomplished with chains 
allows park-unpark 
chains themselves are wait-free
push-detach idiom to wake all
expect multiple waiters on chain to be rare
use XCHG to push and SWAP to detach all 
chain technique is also backwards applicable to TWA itself
Provides futex dispersal to reduce contention on any one chain in kernel
} 

\Invisible{Finally, we believe that replacing the waiting array elements with pointers to chains of waiting threads may have
benefit. Briefly, each long-term waiting thread would have
an on-stack MCS-like queue node that it would push into the
appropriate chain in the waiting array, and then use local
spinning on a field within that node. Notification of longterm waiters causes the chain to be detached via an atomic
SWAP instruction and all the elements are updated to reflect
that they should reevaluate the grant field. In the case of
collisions, waiting threads may need to re-enqueue on the
chain. This design recapitulates much of the Linux kernel ``futex'' mechanism.}

\Invisible{Waiting indicator variations
TKTWA5-TWA5: wait array slots contain update counter; 
TKTWA7-TWA7 : slots contain identity of singular waiter}  

\Invisible{use LFENCE for short-term waiting instead of PAUSE.
LFENCE serves as speculation ``kill'' barrier -- learned via Meltdown/SPECTRE speculative execution attack mitigations}

\section{Empirical Evaluation}
\label{section:Empirical} 

Unless otherwise noted, all data was collected on an Oracle X5-2 system.  
The system has 2 sockets, each populated with 
an Intel Xeon E5-2699 v3 CPU running at 2.30GHz.  Each socket has 18 cores, and each core is 2-way 
hyperthreaded, yielding 72 logical CPUs in total.  The system was running Ubuntu 18.04 with a stock 
Linux version 4.15 kernel, and all software was compiled using the provided GCC version 7.3 toolchain
at optimization level ``-O3''.  
64-bit C++ code was used for all experiments.  
Factory-provided system defaults were used in all cases, and Turbo mode \cite{turbo} was left enabled.  
In all cases default free-range unbound threads were used to provide a realistic evaluation environment.  
 
We used C++ \texttt{std::atomic<>} for low-level atomic operations, but opted for the sake of brevity in explication
to avoid any explicit optimizations related to using relaxed memory order accesses.  
As such, the implementation may be conservatively over-fenced.  
We leave such optimizations, which may be profitable on platforms with weaker memory models, such as ARM, 
for future work.  
All busy-wait loops used the Intel \texttt{PAUSE} instruction for polite waiting.


We use a 128 byte sector size on Intel processors for alignment to avoid 
false sharing.  The unit of coherence is 64 bytes throughout the cache hierarchy, but 128 bytes
is required because of the adjacent cache line prefetch facility where pairs of lines are automatically 
fetched together.

\begin{figure}[h]                                                                    
\includegraphics[width=8.5cm]{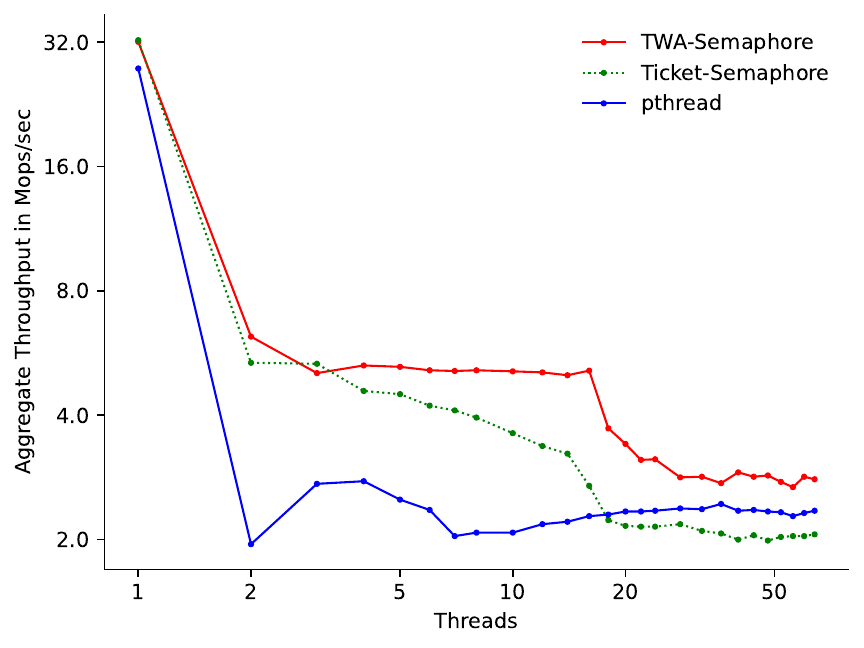} 
\caption{Semaphore Benchmark}                                                   
\label{Figure:semabench}                                                                  
\end{figure}

We modified the \texttt{MutexBench} benchmark \cite{Fissile,dice2020fissile} to replace locks
with semaphores, yielding the \texttt{semabench} benchmark.  
Semabench spawns $T$ concurrent threads. Each thread loops as follows: 
acquire a central shared semaphore S; execute a critical section; release S; execute
a non-critical section. At the end of a 10 second measurement interval the benchmark 
reports the total number of aggregate iterations completed by all the threads. 
We report the median of 11 independent runs in Figure-\ref{Figure:semabench}.  
The critical section advances a shared C++ \texttt{std::mt19937} pseudo-random generator (PRNG)
1 step.  The non-critical section advances a thread-private PRNG instance 1 step.  
To use the semaphore much as a lock, after initializing the semaphore and before any of the 
threads are lauched, we \texttt{post} once to the semaphore.  
For clarity and to convey the maximum amount of information to allow a comparison of the algorithms, 
the $X$-axis is offset to the minimum score and the $Y$-axis is logarithmic.
For reference, we also include the results of the default system user-mode semaphore, \texttt{pthread},
which does not provide FIFO admission.  

\Invisible{To facilitate comparison; visual comparison; to convey maximum information; for clarity; for density; } 

As can be seen in the figure, \emph{Ticket-Semaphore} and \emph{TWA-Semaphore} perform about the same
at low thread counts, with little or no contention.   Performance drops for all the semaphore implementations
as we increase from $1$ to $2$ threads as the benefits from the available parallelism do not yet overcome
the cost of cache coherent communications.  
As we increase the number of threads,
\emph{TWA-Semaphore} outperforms \emph{Ticket-Semaphore}.  Performance drops at about 16 threads
as the scheduler starts to disperse the threads over the NUMA nodes.

while our approach is deterministic
we note that performance under TWA can be influenced by the activities of other unrelated threads and locks
by means of collisions in the shared waiting array, potentially reducing predictability.  Other
shared resources incur the same risk. Examples include (a) competition for occupancy of shared hardware caches
and (b) collisions in the Linux \emph{futex} hash table, where lock addresses map to hash chains of blocked threads. 


\section{Conclusion}

Our approach allows the construction of simple yet scalable semaphores.  
The exhibit state-of-the-art performance at both low and high contention levels, and
are compact, with a highly constrained impact on space.  
They can be easily taylored to use futexes or parking, making them a practical choice
for real-world software.   

\Invisible{Any structure that uses a lock to protect hash chains}  

\bibliography{twa-semaphore.bib}

\section{Appendix : Listings}

\definecolor{forestgreen}{rgb}{0.13, 0.55, 0.13}
\lstset{language=C++}
\lstset{frame=lines}
\lstset{basicstyle=\footnotesize\ttfamily} 
\lstset{commentstyle=\itshape\color{gray}} 
\lstset{commentstyle=\slshape\color{gray}} 
\lstset{commentstyle=\itshape\color{gray}} 
\lstset{keywordstyle=\color{forestgreen}\bfseries} 
\lstset{backgroundcolor=\color{Gray95}} 

\lstset{basicstyle=\fontsize{5.8}{6.5}\selectfont\ttfamily}
\lstset{basicstyle=\fontsize{7}{8}\selectfont\ttfamily}
\lstset{basicstyle=\fontsize{5}{6}\selectfont\ttfamily}
\lstset{basicstyle=\fontsize{6}{7}\selectfont\ttfamily}
\lstset{basicstyle=\fontsize{6.5}{7.5}\selectfont\ttfamily}
\lstset{basicstyle=\fontsize{7}{7.75}\selectfont\ttfamily}

\lstset{numbers=none} 



\newcommand{\IRule}{\smash{\rule[-.2\baselineskip]{.4pt}{\baselineskip}\kern.5em}}


\onecolumn 
\begin{adjustwidth}{1cm}{0pt}

\lstset{caption={Ticket-Semaphore}}
\lstset{label={Listing:Ticket-Semaphore}}
\begin{lstlisting}[language=C++,mathescape=true,escapechar=\%]
// Semaphores implemented with ideas borrowed with ticket locks 
 
struct Semaphore {
 %\IRule% std::atomic<uint64_t> Ticket {0} ;         
 %\IRule% std::atomic<uint64_t> Grant  {0} ;         
} ;

static void Pause() { __asm__ __volatile__ ("rep;nop;") ;} 

static void SemaTake (Semaphore * S) { 
 %\IRule% const auto tx = S%$\rightarrow$%Ticket.fetch_add(1) ; 
 %\IRule% int64_t dx = S%$\rightarrow$%Grant.load() - tx ; 
 %\IRule%  
 %\IRule% // Fast-path uncontended return 
 %\IRule% if (dx > 0) return ; 
 %\IRule%  
 %\IRule% // slow-path contention -- need to wait
 %\IRule% for (;;) { 
 %\IRule%  %\IRule% dx = S%$\rightarrow$%Grant.load() - tx ; 
 %\IRule%  %\IRule% if (dx > 0) return ; 
 %\IRule%  %\IRule% // If desired, we could use a delay proportional in length to dx
 %\IRule%  %\IRule% Pause() ; 
 %\IRule% }
}

static void SemaPost (Semaphore * S) { 
 %\IRule% S%$\rightarrow$%Grant.fetch_add(1) ; 
}



\end{lstlisting}

\lstset{caption={TWA-Semaphore}}
\lstset{label={Listing:TWA-Semaphore}}
\begin{lstlisting}[language=C++,mathescape=true,escapechar=\%]
// Semaphores implemented with ticket locks or TWA

struct Semaphore {
 %\IRule% std::atomic<uint64_t> Ticket {0} ;         
 %\IRule% std::atomic<uint64_t> Grant  {0} ;         
} ;

// Design considerations for the TWAHash() function -- ticket-aware hash function
// The hash function maps Lock address and Ticket value pairs to indices -- ``buckets'' --
// in the long-term waiting array.  
//
// @  Desiderata for the hash function
//    Efficient, low-latency and enables instruction-level parallelism ILP to 
//    the extent possible. 
//    We do NOT generally require a full-strength arbitrary hash function
//    with the usual avalanche property.  
//    Instead, one specialized to the problem is better.  
// @  Index-aware in the sense that we understand that ticket values
//    from a given lock or semaphore increase by 1, which we can use to our advantage.
///   Ideally, as the ticket advances, the index computed by 
//    TWAHash(L,ticket) will "walk" through the full gamut of possible bucket 
//    indices before repeating. 
// @  To reduce coherence traffic and false sharing, adjacent (or numerically proximal) ticket 
//    values should map to indices in the table that reside on different cache lines.
//    A set of waiting threads will have adjacent ticket values but we want them
//    well-dispersed over the buckets to avoid false sharing or even true collisions.
// @  Hash collisions induce extra coherence traffic _and also
//    induce futile wakeups in the buckets, as our wakeup operator
//    activates _all the elements of a bucket and presumes that collisions are rare.
//    We expect in general, however, that only one thread waits on a 
//    given bucket at a given time.  
// @  Cache-friendly and TLB-friendly to the bucket array
// @  Standing in tension to the above, we are also concerned about inter-lock 
//    parallel ticket entrainment, where two different locks or semaphores might move in unison, 
//    inadvertently generate streams of ticket values that continue to collide in the array.  
//    Simple supplementary hashes can mitigate this issue. 
//    Multiplying the "L" lock address component by the "Ticket" component would
//    be ideal but also tends thwart some of our other desiderata, enumerated above. 
//    Broadly, however, the other concerns dominate and have precedence over this 
//    particular design aspect.  
// @  Be aware of automatic stride-based hardware prefetch on some CPUs.
// @  The multiplication constants in the TWAHash() function should be coprime
//    with each other and the length of the array, and also sufficiently
//    large that adjacent ticket values map to indices that map to 
//    different cache lines.  
//    This gives us the optimal "tickets marching through the array" behavior that we 
//    prefer and desire. 
//    The choice of constants is also convolved with the size of the bucket elements.
// @  An interesting variation is to precondition the incoming ticket value : Ticket >>= 1. 
//    This groups threads with numerically adjacent ticket values into pairs.
//    This increases futile wakeups but also yields a pipelined early wakeup effect
//    that can put "near" futile successors a chance to wake early and warmup. 
// @  Another profitable variation divides the ticket value into distinct bit fields.  
//    The upper bits directly identify and select a logical "sub-page" in the waiting array.  
//    The lower bits undergo a hash operation.
//    A given stream of sequential incrementing ticket values will "orbit" within a 
//    sub-page, accessing all elements, and then shift to the next sub-page.  
//    This approach minimizes page transitions for a given stream of ticket values.
//    In turn, that access pattern can act to reduce TLB pressure.  
//    This is somewhat related to "Morton" or Z-order access patterns. 
// @  Gray codes also show promise 

static uint32_t TWAHash (Semaphore * L, uint32_t Ticket) {                               
 %\IRule% return uintptr_t(L) + (Ticket * 17) ; 
}

// Changes to the UpdateSequence field indicate that long-term waiters
// on that particular bucket must should re-check the Grant values to
// see if the waiting condition has been satisfied.  

struct WaitBucket { 
 %\IRule% std::atomic<uint64_t> UpdateSequence {0} ; 
} ; 

// TableSize is a tunable parameter ...
// A sensible policy is to set TableSize to be proportional to the number
// of logical CPUs in the system.
// We force alignment to reduce the number of TLBs underlying the table in
// order to reduce TLB misses.  
// Ideally, we would place the table on large pages.  
// We note that an extremely large table might sound appealing, as that
// design will reduce hash collisions and false wakeups, but a large
// table will also result in cache pollution as the ticket values
// march over the table. 

static WaitBucket * IndexToBucket (uintptr_t A) { 
 %\IRule% constexpr int TableSize = 2048 ; 
 %\IRule% static_assert (TableSize > 0 && (TableSize & (TableSize-1)) == 0) ;
 %\IRule% static WaitBucket WaitSlots [TableSize] __attribute__ ((aligned(4096))) ;
 %\IRule% const int ix = A & (TableSize-1) ; 
 %\IRule% return &WaitSlots [ix] ; 
}

static inline void Pause() { __asm__ __volatile__ ("rep;nop;") ;} 

// LongTermThreshold tunable parameter ...
// Threshold for near-global vs far-local polite waiting
static constexpr int LongTermThreshold = 1 ; 

static void SemaTake (Semaphore * S) { 
 %\IRule% const auto tx = S%$\rightarrow$%Ticket.fetch_add(1) ; 
 %\IRule% int64_t dx = S%$\rightarrow$%Grant.load() - tx ; 
 %\IRule%  
 %\IRule% // Fast-path uncontended return 
 %\IRule% if (dx > 0) return ; 
 %\IRule%  
 %\IRule% // slow-path contention -- need to wait
 %\IRule% const auto s = IndexToBucket(TWAHash(S, tx)) ;
 %\IRule% auto mx = s%$\rightarrow$%UpdateSequence.load() ; 
 %\IRule%  
 %\IRule% for (;;) { 
 %\IRule%  %\IRule% dx = S%$\rightarrow$%Grant.load() - tx ; 
 %\IRule%  %\IRule% if (dx > 0) {
 %\IRule%  %\IRule%  %\IRule% return ; 
 %\IRule%  %\IRule% }
 %\IRule%  %\IRule%  
 %\IRule%  %\IRule% // Short-term waiting -- global waiting on S%$\rightarrow$%Grant
 %\IRule%  %\IRule% // Our thread is near the head of the logical queue of waiting threads.  
 %\IRule%  %\IRule% // It is "proximal" or "near" to acquisition.
 %\IRule%  %\IRule% // As such we switch from semi-local waiting in the WaitArray to global spinning
 %\IRule%  %\IRule% // This strategy yields 2 modes or phases of waiting : near and far 
 %\IRule%  %\IRule% if ((dx + LongTermThreshold) > 0) {
 %\IRule%  %\IRule%  %\IRule% Pause() ; 
 %\IRule%  %\IRule%  %\IRule% continue ; 
 %\IRule%  %\IRule% }
 %\IRule%  %\IRule%  
 %\IRule%  %\IRule% // Long-term distal waiting -- semi-local waiting via the WaitingArray.  
 %\IRule%  %\IRule% // We use indirect waiting on s%$\rightarrow$%UpdateSequence "proxy" indicator
 %\IRule%  %\IRule% // our thread is "far" from the head of the logical queue
 %\IRule%  %\IRule% const auto vx = mx ; 
 %\IRule%  %\IRule% while ((mx = s%$\rightarrow$%UpdateSequence.load()) == vx) {
 %\IRule%  %\IRule%  %\IRule% Pause() ; 
 %\IRule%  %\IRule% }
 %\IRule% }
}

// Optional optimizations ...
// Try to avoid unnecessary long-term notification operations in the common
// uncontended case where there are no waiters.
// Our approach automatically confers the same "fast path" benefits conferred by the
// well-known "benaphore" optimization :  
// https://www.haiku-os.org/legacy-docs/benewsletter/Issue1-26.html#Engineering1-26

static void SemaPost (Semaphore * S) { 
 %\IRule% auto g = S%$\rightarrow$%Grant.fetch_add(1) ; 
 %\IRule% g += LongTermThreshold ; 
 %\IRule%  
 %\IRule% // Optional optimization ...
 %\IRule% // In the case where there are no long-term waiting threads on this specific semaphore, 
 %\IRule% // we would prefer to avoid otherwise unnecessary accesses to the long-term waiting array.   
 %\IRule% // Absent such an optimization, back-to-back sequences of SemaPost() operations would tend to 
 %\IRule% // "march" through the long-term waiting array, increasing cache pressure, with futile accesses.  
 %\IRule% //
 %\IRule% // To this end, we fetch and examine the value of S%$\rightarrow$%Ticket.  
 %\IRule% // If the difference between that observed ticket value and the value we just wrote into
 %\IRule% // grant is less than the LongTermThreshold value then we can safely intuit that there
 %\IRule% // are no long-term waiters that need notification, and can skip the long-term notification step.
 %\IRule% // Note that this approach is conservative and racy.
 %\IRule% // If some other thread calls SemaPost() in the window between the fetch_add(), above, and
 %\IRule% // the fetch of S%$\rightarrow$%Ticket, then we might still perform unnecessary notification, but
 %\IRule% // that is benign in terms of correctness. 
 %\IRule% // That is, the fast-path return does not filter out all unnecessary long-term notifications.
 %\IRule% // 
 %\IRule% // This optimization constitutes a "bet".  
 %\IRule% // The downside is that we incur more coherence traffic as all SemaPost() operations access 
 %\IRule% // both Ticket and Grant fields.   Having said that, the cache line underlying Ticket and Grant is 
 %\IRule% // most likely still in local "M"-state (MESI "modified") at the time of the load() and has 
 %\IRule% // not yet been arbitrated away by concurrent accesses during the fetch_add() - load() window.  
 %\IRule% // The economic "profitability" model for this optimization is thus platform-specific.  
 %\IRule% // The access to Ticket is wasted and futile if we still ultimately need to poke
 %\IRule% // long-duration waiters spinning in the bucket.  
 %\IRule% // 
 %\IRule% // If we fail to take the fast-path return, then access to Ticket was futile
 %\IRule% // and we have incurred a penalty in terms of path length and coherence traffic.
 %\IRule% // So we have a tension and trade-off.
 %\IRule% // The fast-path attempt helps under light load but may incur a performance penalty
 %\IRule% // under contention.  
 %\IRule% //
 %\IRule% // Using the fast-path shortcut return optimization means that the
 %\IRule% // SemaPost() operator must access both Ticket and Grant fields.
 %\IRule% // This precludes or obviates any coherence traffic benefit we might see from 
 %\IRule% // sequestering Ticket and Grant fields on distinct cache lines. 
 %\IRule% int64_t dx = g - S%$\rightarrow$%Ticket.load() ; 
 %\IRule% if (dx >= 0) {
 %\IRule%  %\IRule% return ;        // common fast-path return : short-cut
 %\IRule% }
 %\IRule%  
 %\IRule% // Poke successor-of-successor from long-term mode into short-term mode
 %\IRule% // The immediate successor has already been enabled to enter the critical section.
 %\IRule% // Our approach yields some useful pipeline parallelism as we now act
 %\IRule% // to shift the successor's successor from long-term waiting to short-term waiting mode. 
 %\IRule% const auto s = IndexToBucket (TWAHash(S,g)) ; 
 %\IRule% s%$\rightarrow$%UpdateSequence.fetch_add(1) ; 
}




\end{lstlisting}

\lstset{caption={TWA-Semaphore with waiting chains}}
\lstset{label={Listing:TWA-Semaphore-Chains}}
\begin{lstlisting}[language=C++,mathescape=true,escapechar=\%]
// Semaphores implemented with ticket locks or TWA
// Address-based waiting chains
 
struct Semaphore {
 %\IRule% std::atomic<uint64_t> Ticket {0} ;         
 %\IRule% std::atomic<uint64_t> Grant  {0} ;         
} ;

// Waiting chain services :
// In the original TWA paper, threads busy-waited on an element in the global "WaitArray".
// Those array elements served as proxy modification indicators.  
// We could use exactly the same approach for TWA semaphores.  
//
// But we also desire extend the algorithm to allow spin-the-block spin-then-park waiting via a 
// park-unpark interface so our threads can wait politely in the kernel, surrendering
// the processor to other ready or runnable threads.  
// Specifically, to use park-unpark, we need to map a single waiting array slot to a set
// of waiting threads, as the park-unpark interface needs to be able wake specific threads
// by their identity or associated address.  
// 
// To accomplish this, we change the elements in the WaitArray be pointers to the head
// of a stack of waiting threads instead of a modification counter.  
// Arriving threads push themselves onto the chain with an atomic exchange (SWAP). 
// The list of waiting threads on a given chain is implicit.
// Like CLH locks, and Reciprocating Locks, a thread knows the identity (address) of 
// its immediate successor in the chain, but nothing more.  
// We avoid intrusive linkage -- there is no need for "next" pointers, for instance
// Once a thread has pushed its "WaitElement" onto the chain, it then busy-waits (or parks)
// on a the "Gate" field within that WaitElement.  
// Our design intentionally allows WaitElements to be allocated on-stack, which much simplifies 
// the design and removes memory management and lifecycle entanglements.
// 
// To notify and wake threads, we detach the _entire chain with an exchange(nullptr) operation
// and activate the 1st thread in the chain.  That thread, in turn, activates its successor,
// and so on, in a systolic fashion, to make sure all the detached elements are made awake 
// so they can recheck their respective conditions of interest.
// Hash collision can result in multiple threads emplaced on a given chain, and thus 
// our waiting design admits spurious wakeups, so either the waiting primitives or their caller 
// must reevaluate the condition before determining the wait is satisfied.
// 
// The key operators on the concurrent stack are push-one and detach-all. 
// We observe that a "pop-stack" structure, which supports push and detach-all operations,
// is well-known : https://mathscinet.ams.org/mathscinet/relay-station?mr=0624050. 
// We implement our waiting chains as a specialized  _concurrent pop-stack. 
// We use a wait-one wake-all notification policy.
//
// In lieu of a full park-unpark interface, we could also use linux kernel "Futex" waiting. 
// on the WaitElement "Gate" fields. 
//
// We note that our waiting chain design is applicable and could be used in TWA itself
// as an improvement to that algorithm.
//
// Our waiting chain construction converts global spinning (where multiple threads might
// busy-wait on one location) into local 1-to-1 spinning where at most one thread waits 
// on a given location.  
// In turn, this transformation makes waiting amenable to blocking primitives such
// as park-park facilities, or linux kernel futex operators. 
//
// Given that we always wake _all threads on the chain, the mechanism exhibits quadratic 
// wakeup  complexity.  We expect, however, that the number of threads even found 
// concurrently on a chain to be extremely small. 


// Harmonize C++ std::atomic API with conventions in the academic literature :
// The following lets us write compare-and-swap cas() 
template <typename T> class Atomic : public std::atomic<T> {
 public :
 %\IRule% T cas (T cmp, T set) { this%$\rightarrow$%compare_exchange_strong (cmp, set) ; return cmp ; } 
 %\IRule% T operator=(T rhs)   { this%$\rightarrow$%store(rhs); return rhs; }  
} ;  

// We impose alignment on WaitElement to reduce on-stack false sharing
// We use 128 instead of 64 to avoid adjacent sector prefetch issues on x86.  
// WaitElement instances are thread-specific.
// That is, at most one thread is associated with or waiting on a WaitElement
// at any given time.

struct alignas(128) WaitElement {
 %\IRule% std::atomic<int> Gate {0} ;       // made ready -- notified
 %\IRule% uintptr_t who ;                   // thread identity for park-unpark  
} ; 
 
// The Waiting table is an array of WaitChain elements -- implemented as a flat hashtable.
// These WaitChain instances are the "buckets" (slots or cells) in the hashtable.  
// In turn, each WaitChain holds a pointer to the front of a list of WaitElements.  
// Individual threads wait on those WaitElements. 
// We hash into the Waiting table via an address to select a specific WaitChain instance.
// As noted, those slots contain a pointer to a list of waiting threads that are currently 
// monitoring addresses that happen to map to that table index. 

struct WaitChain {              // AKA : WaitingList or WaitingThreads or WaitBucket                  
 %\IRule% Atomic<WaitElement *> Chain {nullptr} ; 
} ; 

// TableSize is a tunable parameter ...
// A sensible policy would be to set TableSize to be proportional to the number
// of logical CPUs in the system.
// We force alignment to reduce the number of virtual pages underlying the 
// table in order to reduce TLB (translation lookaside buffer) misses. 
// Ideally, we would place the table on large pages, if available.
// We note that an excessively large table might sound appealing, as that
// design will reduce hash collisions and false wakeups, but a large
// table will also result in cache pollution as the ticket values
// march over the table. 
// Empirically, 4096 elements is a reasonable choice for most systems.  

static WaitChain * KeyToChain (uintptr_t A) { 
 %\IRule% // If desired, we could also apply an optional secondary hash to A
 %\IRule% constexpr int TableSize = 4096 ;
 %\IRule% static_assert (TableSize > 0 && (TableSize & (TableSize-1)) == 0) ;
 %\IRule% static WaitChain WaitSlots [TableSize] __attribute__ ((aligned(4096))) ;
 %\IRule% const int ix = A & (TableSize-1) ; 
 %\IRule% return &WaitSlots [ix] ; 
}

static void Poke (WaitElement * e) { 
 %\IRule% if (e == nullptr) return ; 
 %\IRule% auto who = e%$\rightarrow$%Who ; 
 %\IRule% assert (e%$\rightarrow$%Gate == 0) ; 
 %\IRule% e%$\rightarrow$%Gate.store(1, std::memory_order_release); 
 %\IRule% // Once we set e%$\rightarrow$%Gate=1, it is possible e will fall immediately out of scope.
 %\IRule% // That is, once we set e%$\rightarrow$%Gate=1, it is no longer safe to access e.  
 %\IRule% // It is safe, however, to unpark a stale thread reference.
 %\IRule% Unpark (who) ; 
}

static void AddressSignal (uintptr_t A) { 
 %\IRule% const auto slot = KeyToChain (A) ; 
 %\IRule% Poke (slot%$\rightarrow$%Chain.exchange(nullptr)) ; 
}

// The "polite" form avoids mutating the chain pointer if it is already empty,
// reducing coherence traffic if there are no waiting threads.

static void AddressSignalPolite (uintptr_t A) { 
 %\IRule% const auto slot = KeyToChain (A) ; 
 %\IRule% if (slot%$\rightarrow$%Chain.load() %$\neq$% nullptr) { 
 %\IRule%  %\IRule% Poke (slot%$\rightarrow$%Chain.exchange(nullptr)) ; 
 %\IRule% } 
}


static uintptr_t AddressWaitUntil (uintptr_t A, std::invocable auto && Satisfied) { 
 %\IRule% // Optional optimization :
 %\IRule% // Before our waiter goes to the expense of becoming visible on the chain
 %\IRule% // we re-evaluate the condition to see if we skip the more expensive steps.
 %\IRule% // Could also implement a bounded spinning phase here, which
 %\IRule% // is tantamount to global spinning.  
 %\IRule% uintptr_t v = Satisfied() ; 
 %\IRule% if (v %$\neq$% 0) return v ; 
 %\IRule%  
 %\IRule% // Switch to a more polite waiting strategy -- local spinning
 %\IRule% const auto s = KeyToChain (A) ; 
 %\IRule% for (;;) { 
 %\IRule%  %\IRule% // Optional optimization but reduces coherence traffic on the chain
 %\IRule%  %\IRule% v = Satisfied() ; 
 %\IRule%  %\IRule% if (v %$\neq$% 0) return v ; 
 %\IRule%  %\IRule%  
 %\IRule%  %\IRule% // Emplace candidate element E on the wait chain -- start monitoring A.  
 %\IRule%  %\IRule% // At this point our thread becomes a "visible" waiter and has "published" E  
 %\IRule%  %\IRule% // which expresses our interest in updates to A.  
 %\IRule%  %\IRule% // We now wait via E instead of busy-waiting directly on A.  
 %\IRule%  %\IRule% // Pass thread identity from waiter to wakee via the "who" field
 %\IRule%  %\IRule% WaitElement E {} ;  
 %\IRule%  %\IRule% E.who = Self() ; 
 %\IRule%  %\IRule% const auto prv = s%$\rightarrow$%Chain.exchange (&E) ; 
 %\IRule%  %\IRule% assert (prv %$\neq$% &E) ; 
 %\IRule%  %\IRule%  
 %\IRule%  %\IRule% // Recheck the condition -- ratify and confirm we still need to wait
 %\IRule%  %\IRule% // This closes a race condition where a thread in AddressWaitUntil() 
 %\IRule%  %\IRule% // concurrently races against one in AddressSignal().  
 %\IRule%  %\IRule% // We recheck to avoid lost or missed wakeup pathologies. 
 %\IRule%  %\IRule% // 
 %\IRule%  %\IRule% // Dekker Duality and "pivot" : 
 %\IRule%  %\IRule% // Wait : ST Chain; LD Condition 
 %\IRule%  %\IRule% // Post : ST Condition; LD Chain
 %\IRule%  %\IRule%  
 %\IRule%  %\IRule% v = Satisfied() ; 
 %\IRule%  %\IRule% if (v %$\neq$% 0) {
 %\IRule%  %\IRule%  %\IRule% // We mis-queued ourself -- need to recover
 %\IRule%  %\IRule%  %\IRule% // We expect this path to be rarely taken
 %\IRule%  %\IRule%  %\IRule% // Abort the waiting operation and get E off the chain.  
 %\IRule%  %\IRule%  %\IRule% // While the condition of interest is satisfied, we can't yet return
 %\IRule%  %\IRule%  %\IRule% // until E is off the chain and our successors have been notified.
 %\IRule%  %\IRule%  %\IRule% // We must ensure E is off-list and thus privatized/localized before we return. 
 %\IRule%  %\IRule%  %\IRule% //
 %\IRule%  %\IRule%  %\IRule% // We apply a number of optional optimizations to try avoid full
 %\IRule%  %\IRule%  %\IRule% // flush of the chain.  
 %\IRule%  %\IRule%  %\IRule% // 
 %\IRule%  %\IRule%  %\IRule% // optional optimization : try to recover with atomic compare-and-swap (CAS).                            
 %\IRule%  %\IRule%  %\IRule% // See if we can simply "undo" the push, above. 
 %\IRule%  %\IRule%  %\IRule% // This optimization succeeds if no other threads have pushed in the exchange-CAS window.  
 %\IRule%  %\IRule%  %\IRule% // This is safe and immune to ABA pathologies as no other thread can emplace &E.
 %\IRule%  %\IRule%  %\IRule% // Threads only enqueue themselves, not others.                                   
 %\IRule%  %\IRule%  %\IRule% // and the chain suffix is stable if the head refers to our same element, E.    
 %\IRule%  %\IRule%  %\IRule% const auto k = s%$\rightarrow$%Chain.cas(&E, prv) ; 
 %\IRule%  %\IRule%  %\IRule% if (k == &E) {
 %\IRule%  %\IRule%  %\IRule%  %\IRule% assert (E.Gate == 0) ; 
 %\IRule%  %\IRule%  %\IRule%  %\IRule% return v ;  
 %\IRule%  %\IRule%  %\IRule% }
 %\IRule%  %\IRule%  %\IRule% 
 %\IRule%  %\IRule%  %\IRule% // Optional optimization for common case 
 %\IRule%  %\IRule%  %\IRule% if (k == nullptr) { 
 %\IRule%  %\IRule%  %\IRule%  %\IRule% Poke (prv); 
 %\IRule%  %\IRule%  %\IRule%  %\IRule% while (E.Gate == 0) { 
 %\IRule%  %\IRule%  %\IRule%  %\IRule%  %\IRule% Park() ; 
 %\IRule%  %\IRule%  %\IRule%  %\IRule% }
 %\IRule%  %\IRule%  %\IRule%  %\IRule% return v ; 
 %\IRule%  %\IRule%  %\IRule% }
 %\IRule%  %\IRule%  %\IRule%  
 %\IRule%  %\IRule%  %\IRule% // optional optimization
 %\IRule%  %\IRule%  %\IRule% // If E is found already off-list then we can dispense with and avoid
 %\IRule%  %\IRule%  %\IRule% // the full flush
 %\IRule%  %\IRule%  %\IRule% if (E.Gate %$\neq$% 0) { 
 %\IRule%  %\IRule%  %\IRule%  %\IRule% Poke (prv) ; 
 %\IRule%  %\IRule%  %\IRule%  %\IRule% return v ; 
 %\IRule%  %\IRule%  %\IRule% }
 %\IRule%  %\IRule%  %\IRule%  
 %\IRule%  %\IRule%  %\IRule% // We need to use a full chain flush ...
 %\IRule%  %\IRule%  %\IRule% // Eject and wake all the elements
 %\IRule%  %\IRule%  %\IRule% // QoI : we can wake the chain elements in any order, but it's more polite to wakeup 
 %\IRule%  %\IRule%  %\IRule% // the older elements in the suffix (identified via prv) before the newer elements 
 %\IRule%  %\IRule%  %\IRule% // in the just detached prefix.
 %\IRule%  %\IRule%  %\IRule% // 
 %\IRule%  %\IRule%  %\IRule% // prv points to the suffix
 %\IRule%  %\IRule%  %\IRule% // s%$\rightarrow$%Chain points to the prefix
 %\IRule%  %\IRule%  %\IRule% // It is safe to wake the prefix and suffix segments in either order.  
 %\IRule%  %\IRule%  %\IRule% // But if we use "Poke(Prv); Poke(s%$\rightarrow$%Chain.exchange(null))" then the suffix elements
 %\IRule%  %\IRule%  %\IRule% // may race our thread and simply re-queue or migrate back onto s%$\rightarrow$%Chain, 
 %\IRule%  %\IRule%  %\IRule% // before we execute the exchange(), resulting in futile concurrent chain activity.
 %\IRule%  %\IRule%  %\IRule% // This is benign but may impact performance. 
 %\IRule%  %\IRule%  %\IRule% const auto Prefix = s%$\rightarrow$%Chain.exchange (nullptr) ; 
 %\IRule%  %\IRule%  %\IRule% assert ((prv %$\neq$% Prefix) || (prv == nullptr && Prefix == nullptr)) ; 
 %\IRule%  %\IRule%  %\IRule% Poke (prv) ; 
 %\IRule%  %\IRule%  %\IRule% Poke (Prefix) ; 
 %\IRule%  %\IRule%  %\IRule% while (E.Gate == 0) { 
 %\IRule%  %\IRule%  %\IRule%  %\IRule% Park () ; 
 %\IRule%  %\IRule%  %\IRule% }
 %\IRule%  %\IRule%  %\IRule% return v ;  
 %\IRule%  %\IRule% }
 %\IRule%  %\IRule%  
 %\IRule%  %\IRule% // We are properly enqueued on the chain and can now just wait politely
 %\IRule%  %\IRule% // We expect this to be the normal dominant case
 %\IRule%  %\IRule% while (E.Gate == 0) {
 %\IRule%  %\IRule%  %\IRule% Park () ; 
 %\IRule%  %\IRule% }
 %\IRule%  %\IRule%  
 %\IRule%  %\IRule% // propagate the systolic wakeup through the remainder of the stack, if any ...
 %\IRule%  %\IRule% Poke (prv) ; 
 %\IRule%  %\IRule%  
 %\IRule%  %\IRule% // Cases :
 %\IRule%  %\IRule% // @  We were correctly signaled
 %\IRule%  %\IRule% // @  We were purged by a flush operation 
 %\IRule%  %\IRule% // @  We were purged by a hash collision
 %\IRule%  %\IRule% // 
 %\IRule%  %\IRule% // Our thread is now off the chain and it is safe for E to fall out of scope.
 %\IRule%  %\IRule% // There will be no subsequent concurrent accesses to E.  
 %\IRule%  %\IRule% // E is privatized and localized -- detached. 
 %\IRule%  %\IRule% // We need to reevaluate the condition as we might have been ejected from the 
 %\IRule%  %\IRule% // chain via false or spurious wakeups arising from hash collisions.
 %\IRule%  %\IRule% // If necessary, we will re-push ourselves on the stack and resume waiting.
 %\IRule%  %\IRule% // False wakeups are benign and a performance concern, not a correctness issue. 
 %\IRule%  %\IRule% // As the ticket values and their hash (TWAHash) disperse waiting over the 
 %\IRule%  %\IRule% // set of chains, we expect false wakeups to be rare. 
 %\IRule%  %\IRule% // And specifically we expect that having more than one element on a given chain
 %\IRule%  %\IRule% // to be a rare condition.  
 %\IRule%  %\IRule% // A lazy implementation might return here and let the higher level 
 %\IRule%  %\IRule% // decide whether to re-evaluate the condition and then loop or not.
 %\IRule%  %\IRule% // Our implementation happens to be strict and persistent.  
 %\IRule% }
}
 
// In normal "classic" usage, we might use AddressWaitUntil() and AddressSignal() as follows.
// Say a global shared variable X is initially 0 and thread T1 desires to wait until X becomes 1. 
// Thread T2, at some point in the future, eventually sets X to 1.  
// T1 would execute : AddressWaitUntil (&X, [&]{ return X == 1;}) ; 
// T2 would execute : X=1; AddressSignal(&X) 
// 
// For ticket-based waiting, however, we can be more clever and instead of using an simple address,
// we can compose the Ticket value into an abstract "key" which we use in lieu of the
// address of a shared global variable.  This confers a distinct performance advantage,
// as AddressSignal(), absent any hash collisions, will selectively wake up and notify
// just the single waiting thread that is interested in that particular ticket value.
// We use TWAHash() to map the address of the semaphore instance and ticket value to 
// such abstract keys.  
//
// As noted, these "synthetic" ticket-based keys disperse waiting over the slots in the table and
// act to keep the chains short, which improves performance and reduces the rate of 
// futile wakeups that would otherwise occur if we just forced all waiting to 
// use a single address.

static uint32_t TWAHash (Semaphore * L, uint32_t Ticket) {                               
 %\IRule% return uintptr_t(L) + (Ticket * 17) ; 
}

static void SemaTake (Semaphore * S) { 
 %\IRule% const auto tx = S%$\rightarrow$%Ticket.fetch_add(1) ; 
 %\IRule% // dx represents the "distance" or surplus 
 %\IRule% int64_t dx = S%$\rightarrow$%Grant.load() - tx ; 
 %\IRule%  
 %\IRule% // Fast-path uncontended return 
 %\IRule% if (dx > 0) return ; 
 %\IRule%  
 %\IRule% // slow-path contention -- need to wait
 %\IRule% // Compose the address of the semaphore and the ticket value into a key
 %\IRule% // A simple XOR or "+" of the semaphore's address and the ticket value suffices.
 %\IRule% auto Key = TWAHash(S, tx) ; 
 %\IRule% AddressWaitUntil (Key, [&]{ return S%$\rightarrow$%Grant.load() > tx ; }) ; 
 %\IRule% assert (s%$\rightarrow$%Grant.load() > tx) ; 
}

static void SemaPost (Semaphore * S) { 
 %\IRule% auto g = S%$\rightarrow$%Grant.fetch_add(1) ; 
 %\IRule% auto Key = TWAHash (S, g) ; 
 %\IRule% AddressSignal (Key) ; 
}



\end{lstlisting}

\lstset{caption={TWA-Semaphore with MONITOR-MWAIT inspired waiting chains}}
\lstset{label={Listing:TWA-Semaphore-Chains-Monitor}}
\lstinputlisting[language=C++,mathescape=true,escapechar=\%]{TWASemaphore-ByAddress-Monitor-Listing.cc-fx}

\lstset{caption={TWA-Semaphore implemeneted with \texttt{LocationWait()} primitive}}
\lstset{label={Listing:TWA-Semaphore-Chains-LocationWait}}
\begin{lstlisting}[language=C++,mathescape=true,escapechar=\%]
// @  Uses LocationWait() and LocationWakeAll() for long-term 
//    "polite" waiting 
//
// @  LocationWait() is implemented via Concurrent pop-stack
//    push-one; detach-all
//    Quadratic wakeup complexity
//    Presume use case where most chains have at most one element 
//    Collisions and multiple threads waiting on the same location is rare 
//    Related to Reciprocating locks arrival stack
//
// @  Key obvious enabling observation ...
//    A thread can wait only only one condition at a given time
// 
// @  Provides "Proxy" indirect waiting 
// 
// @  Dekker duality pivot
//    Complementary code in WakeAll() and Wait() 
//    *  WakeAll : ST Cond; Swap Slot (null)   
//    *  Wait    : Swap Slot(E) ; LD Cond     
//       Use "LD Cond" to ratify and confirm to close race against concurrent WakeAll 
// 
// @  Unrolled access pattern imposed by the state machine 
//    (1)  check Cond
//    (2)  Emplace E on channel then return immediately
//    (3)  check Cond -- ratification to close race vs signal
//    (4)  Wait 
//
//    The resultant loop structure is subtle as we alternate iterations, 
//    emplacing and waiting, and use the loop to check or recheck the condition
//    to close races, but the result is extremely easy to use. 
//
// @  Key enabling observation for simplification optimization :
//    *   If ratification phase fails then we just exit the loop but allow 
//        element E potentially linger on chain 
//        E has been abandoned and is stranded on the chain.
//        We allow this condition to make the API usage more simple.  
//    *   As we won't get control if the developer breaks out of the 
//        loop when E is emplaced, we deal with E in a lazy fashion
//        and defer and avoid the flush-and-wait operation until
//        the next time we need to use E -- the next waiting episode -- 
//        or until the thread dies. 
//    *   Defer strategy requires E in TLS instead of on-stack allocation !
//    *   Requires thread DTOR in WaitEvent for cleanup check 
//        Beware that DTORs in TLS triggers runtime compiler-emitted registration guards
//        WaitElement then becomes non-trivial constructable/destructable 
//        The runtime guards, on the 1st access by a thread to WaitEvent,
//        will call _cxa_thread_atexit() to register a thread-shut destructor callback. 
//
// @  Location triage for WaitElement E :
//    *   free-floating and not on chain -- ready to be emplaced 
//    *   emplaced on correct chain and ready to wait
//    *   because of deferred flush resides on wrong chain -- detritus that requires flushing
//
// @  Additional pipelining optional optimization
//    Allocate array of 2 WaitElement in TLS
//    Decouple flush phase and wait-for-flush phase
//    Alternate between [0] and [1] 
 

struct TWASemaphoreV3 {
 %\IRule%  
 %\IRule% struct WaitSlot ; 
 %\IRule%  
 %\IRule% struct WaitElement {
 %\IRule%  %\IRule% std::atomic <int> Gate alignas(128) {0} ; 
 %\IRule%  %\IRule% // Where is accessed _only by the owner -- no concurrency
 %\IRule%  %\IRule% // It tells us what WaitSlot "channel" E resides on 
 %\IRule%  %\IRule% WaitSlot * Where = nullptr ;                       
 %\IRule%  %\IRule% std::atomic <WaitElement *> Succ {nullptr} ; 
 %\IRule%  %\IRule%  
 %\IRule%  %\IRule% WaitElement()  {} 
 %\IRule%  %\IRule% ~WaitElement() { 
 %\IRule%  %\IRule%  %\IRule% if (Where %$\neq$% nullptr && Gate.load() == 0) { 
 %\IRule%  %\IRule%  %\IRule%  %\IRule% Poke (Where%$\rightarrow$%Chain.exchange (nullptr)) ; 
 %\IRule%  %\IRule%  %\IRule%  %\IRule% while (Gate.load() == 0) {
 %\IRule%  %\IRule%  %\IRule%  %\IRule%  %\IRule% Pause() ; 
 %\IRule%  %\IRule%  %\IRule%  %\IRule% }
 %\IRule%  %\IRule%  %\IRule% }
 %\IRule%  %\IRule% } 
 %\IRule% } ; 
 %\IRule%  
 %\IRule% struct WaitSlot {         
 %\IRule%  %\IRule% Atomic <WaitElement *> Chain {nullptr} ; 
 %\IRule% } ;
 %\IRule%  
 %\IRule% static inline thread_local WaitElement E {} ; 
 %\IRule%  
 %\IRule% // For the classic concurrent pop-stack, the Poke() operator only needs to 
 %\IRule% // detach and poke the head and the systolic mechanism takes care
 %\IRule% // of propagating the notification thru the chain.
 %\IRule% // But as elements can now be submerged on the chain and transiently abandoned, 
 %\IRule% // we dont always have a waiting thread ready to propagate, 
 %\IRule% // so Poke() needs to walk the chain.
 %\IRule% // When we set k%$\rightarrow$%Gate = 1 we can not rely on k's thread to be waiting
 %\IRule% // and propagate the notifiction to k's successors.  
 %\IRule% // That is, when we poke "k" there is not necessary a thread waiting
 %\IRule% // on k to continue propagation of the wakeups.  
 %\IRule% // If the chain points to A and A's successor is B and A is orphan
 %\IRule% // and the chain is notified, then B can be occluded by A resulting
 %\IRule% // in progress failure.  
 %\IRule% // So instead, we need to walk the chain in Poke() 
 %\IRule% //
 %\IRule% // To make the stack list walkable for Poke() we use a LD-CAS loop but 
 %\IRule% // abandon and forego the constant-time arrival property.
 %\IRule% // While not desirable, we don't really care about order as we are using a stack.  
 %\IRule% // And the technique is lock-free.
 %\IRule%  
 %\IRule% static void Poke (WaitElement * e) {
 %\IRule%  %\IRule% while (e %$\neq$% nullptr) { 
 %\IRule%  %\IRule%  %\IRule% auto k = e ; 
 %\IRule%  %\IRule%  %\IRule% e = k%$\rightarrow$%Succ.load() ;
 %\IRule%  %\IRule%  %\IRule% assert (e %$\neq$% k) ; 
 %\IRule%  %\IRule%  %\IRule% assert (k%$\rightarrow$%Gate.load() %$\neq$% 0) ; 
 %\IRule%  %\IRule%  %\IRule% k%$\rightarrow$%Gate.store (1, std::memory_order_release) ; 
 %\IRule%  %\IRule% }
 %\IRule% }
 %\IRule%  
 %\IRule% static void LocationWait (WaitSlot * s) {
 %\IRule%  %\IRule% // Advance thread-local state machine based on E.Gate and E.Where
 %\IRule%  %\IRule% // We alternate emplace and wait phases 
 %\IRule%  %\IRule% assert (s %$\neq$% nullptr) ;
 %\IRule%  %\IRule% auto where = E.Where ; 
 %\IRule%  %\IRule% if (where == s) { 
 %\IRule%  %\IRule%  %\IRule% // previously emplaced on correct chain 
 %\IRule%  %\IRule%  %\IRule% while (E.Gate.load(std::memory_order_acquire) == 0) { 
 %\IRule%  %\IRule%  %\IRule%  %\IRule% Pause() ; 
 %\IRule%  %\IRule%  %\IRule% }
 %\IRule%  %\IRule%  %\IRule%  
 %\IRule%  %\IRule%  %\IRule% // Clearing E.Succ isn't necessary but is good hygiene
 %\IRule%  %\IRule%  %\IRule% E.Succ.store (nullptr, std::memory_order_release) ; 
 %\IRule%  %\IRule%  %\IRule% E.Where = nullptr ; 
 %\IRule%  %\IRule%  %\IRule% return ; 
 %\IRule%  %\IRule% }
 %\IRule%  %\IRule%  
 %\IRule%  %\IRule% if (where %$\neq$% nullptr) {
 %\IRule%  %\IRule%  %\IRule% // We found E has residual residency on the wrong chain
 %\IRule%  %\IRule%  %\IRule% // E was left as orphan so we perform deferred recovery and cleanup.
 %\IRule%  %\IRule%  %\IRule% // Need to recover and extricate E before we emplace it on the correct chain
 %\IRule%  %\IRule%  %\IRule% // This is a consequence of deferred lazy removal 
 %\IRule%  %\IRule%  %\IRule% // We can not reuse E until we get if off its current chain
 %\IRule%  %\IRule%  %\IRule% if (E.Gate.load(std::memory_order_acquire) == 0) {
 %\IRule%  %\IRule%  %\IRule%  %\IRule% // flush chain and wait for E to become enabled 
 %\IRule%  %\IRule%  %\IRule%  %\IRule% Poke (where%$\rightarrow$%Chain.exchange (nullptr)) ; 
 %\IRule%  %\IRule%  %\IRule%  %\IRule% while (E.Gate.load(std::memory_order_acquire) == 0) { 
 %\IRule%  %\IRule%  %\IRule%  %\IRule%  %\IRule% Pause() ; 
 %\IRule%  %\IRule%  %\IRule%  %\IRule% }
 %\IRule%  %\IRule%  %\IRule% }
 %\IRule%  %\IRule%  %\IRule%  
 %\IRule%  %\IRule%  %\IRule% // The following two stores are not strictly necessary but
 %\IRule%  %\IRule%  %\IRule% // are useful for assert()s and diagnostic hygiene
 %\IRule%  %\IRule%  %\IRule% E.Where = nullptr ; 
 %\IRule%  %\IRule%  %\IRule% E.Succ.store (nullptr, std::memory_order_release) ; 
 %\IRule%  %\IRule% }
 %\IRule%  %\IRule%  
 %\IRule%  %\IRule% // E currently resides on no chain and is privatized-localized 
 %\IRule%  %\IRule% // Not reachable by any concurrent actions 
 %\IRule%  %\IRule% // Emplace E on chain s
 %\IRule%  %\IRule% E.Where = s ; 
 %\IRule%  %\IRule% E.Gate.store (0, std::memory_order_release) ; 
 %\IRule%  %\IRule%  
 %\IRule%  %\IRule% // Lock-free LD-CAS loop .  
 %\IRule%  %\IRule% // Consider : optional optimistic optimization
 %\IRule%  %\IRule% // as we expect slots to be mostly empty, we could start out 
 %\IRule%  %\IRule% // with optimistic cas() attempt that guesses that the s%$\rightarrow$%Chain is null.
 %\IRule%  %\IRule% E.Succ.store (nullptr, std::memory_order_release) 
 %\IRule%  %\IRule% auto succ = s%$\rightarrow$%Chain.cas (nullptr, &E) ; 
 %\IRule%  %\IRule% if (succ == nullptr) return 
 %\IRule%  %\IRule% for (;;) {
 %\IRule%  %\IRule%  %\IRule% assert (succ %$\neq$% &E) ; 
 %\IRule%  %\IRule%  %\IRule% // Tenative store in anticipation of successful cas() 
 %\IRule%  %\IRule%  %\IRule% E.Succ.store (succ, std::memory_order_release) ; 
 %\IRule%  %\IRule%  %\IRule% auto v = s%$\rightarrow$%Chain.cas (succ, &E) ; 
 %\IRule%  %\IRule%  %\IRule% if (v == succ) [[likely]] break ; 
 %\IRule%  %\IRule%  %\IRule% succ = v ; 
 %\IRule%  %\IRule%  %\IRule% // Inopportune interleaving -- we encountered concurrent interference
 %\IRule%  %\IRule%  %\IRule% // we raced and lost, but some other thread made progress
 %\IRule%  %\IRule%  %\IRule% // Some other thread updated s%$\rightarrow$%Chain in the LD-CAS window
 %\IRule%  %\IRule%  %\IRule% // Just retry 
 %\IRule%  %\IRule% }
 %\IRule%  %\IRule%  
 %\IRule%  %\IRule% // control returns to the loop to perform ratify check
 %\IRule%  %\IRule% // condition -- location of interest 
 %\IRule%  %\IRule% // close race against concurrent notification 
 %\IRule%  %\IRule% // 
 %\IRule%  %\IRule% // On the _next call to LocationWait(), we will actually wait
 %\IRule%  %\IRule% // From the callers perspective this appears to be a spurious
 %\IRule%  %\IRule% // return from LocationWait() but we do so intentionally to
 %\IRule%  %\IRule% // reevaluate the condition before the next iteration,
 %\IRule%  %\IRule% // which will actually wait. 
 %\IRule% }
 %\IRule%  
 %\IRule% // Consider : polite LD-if-exchange form
 %\IRule% // Need some of type ST-LD barrier between the prior store
 %\IRule% // that changes the waited-upon condition and the LD of s%$\rightarrow$%Chain
 %\IRule% static void LocationWakeAll (WaitSlot * s) { 
 %\IRule%  %\IRule% assert (s %$\neq$% nullptr) ; 
 %\IRule%  %\IRule% Poke (s%$\rightarrow$%Chain.exchange (nullptr)) ; 
 %\IRule% } 
 %\IRule%  
 %\IRule% [[gnu::pure]] static WaitSlot * IndexToBucket (uintptr_t A) { 
 %\IRule%  %\IRule% constexpr int TableSize = 4096 ; 
 %\IRule%  %\IRule% static_assert (TableSize > 0 && (TableSize & (TableSize-1)) == 0) ;
 %\IRule%  %\IRule% static WaitSlot WaitSlots [TableSize] __attribute__ ((aligned(4096))) ;
 %\IRule%  %\IRule% const int ix = A & (TableSize-1) ; 
 %\IRule%  %\IRule% return &WaitSlots [ix] ; 
 %\IRule% }
 %\IRule%  
 %\IRule% [[gnu::pure]] uint32_t TWAHash (uint32_t Ticket) {                               
 %\IRule%  %\IRule% return uint32_t(uintptr_t(this)) + (Ticket * 17) ; 
 %\IRule% }
 %\IRule%  
 %\IRule% [[gnu::pure]] static uint32_t Mix32A (uint32_t v) {
 %\IRule%  %\IRule% static const int32_t Mix32KA = 0x9abe94e3 ; 
 %\IRule%  %\IRule% v = (v ^ (v >> 16)) * Mix32KA ; 
 %\IRule%  %\IRule% v = (v ^ (v >> 16)) * Mix32KA ; 
 %\IRule%  %\IRule% return v ^ (v >> 16) ; 
 %\IRule% }
 %\IRule%  
 %\IRule% // General purpose address-based forms ...
 %\IRule% // These use a generic hash function based on Mix32A() instead of
 %\IRule% // the intentionally designed TWAHash() operator that works well
 %\IRule% // with ticket values.  
 %\IRule% // 
 %\IRule% // A trivial but extremely useful optimization for the general forms
 %\IRule% // is to implement a singleton thread_local KV cache that maps the
 %\IRule% // Address A to its corresponding slot pointer.  
 %\IRule%  
 %\IRule% static WaitSlot * Lookup (uintptr_t A) { 
 %\IRule%  %\IRule% static constinit thread_local WaitSlot * CacheSlot = nullptr ; 
 %\IRule%  %\IRule% static constinit thread_local uintptr_t CacheKey = 0 ; 
 %\IRule%  %\IRule% auto s = CacheSlot ; 
 %\IRule%  %\IRule% if (A == 0) A = 0xD1CE ; 
 %\IRule%  %\IRule% if (A %$\neq$% CacheKey) {
 %\IRule%  %\IRule%  %\IRule% s = IndexToBucket (Mix32A(uint32_t(A))) ; 
 %\IRule%  %\IRule%  %\IRule% CacheSlot = s ; 
 %\IRule%  %\IRule%  %\IRule% CacheKey  = A ; 
 %\IRule%  %\IRule% }
 %\IRule%  %\IRule% assert (s %$\neq$% nullptr) ; 
 %\IRule%  %\IRule% return s ; 
 %\IRule% }
 %\IRule%  
 %\IRule% static void LocationWait (uintptr_t A) { 
 %\IRule%  %\IRule% LocationWait (Lookup(A)) ; 
 %\IRule% }
 %\IRule%  
 %\IRule% static void LocationWakeAll (uintptr_t A) { 
 %\IRule%  %\IRule% LocationWakeAll (Lookup(A)) ;
 %\IRule% }
 %\IRule%  
 %\IRule% TWASemaphoreV3 (int N) { 
 %\IRule%  %\IRule% assert (N >= 0) ; 
 %\IRule%  %\IRule% Grant.store (N) ; 
 %\IRule% }
 %\IRule%  
 public :
 %\IRule% std::atomic<uint64_t> Ticket {0} ;         
 %\IRule% std::atomic<uint64_t> Grant  {0} ;         
 %\IRule%  
 %\IRule% void SemaTake () { 
 %\IRule%  %\IRule% const auto tx = Ticket.fetch_add(1) ; 
 %\IRule%  %\IRule% int64_t dx = Grant.load() - tx ; 
 %\IRule%  %\IRule%  
 %\IRule%  %\IRule% // Fast-path uncontended return 
 %\IRule%  %\IRule% if (dx > 0) return ; 
 %\IRule%  %\IRule%  
 %\IRule%  %\IRule% // slow-path contention -- need to wait
 %\IRule%  %\IRule% const auto s = IndexToBucket(TWAHash(tx)) ;
 %\IRule%  %\IRule%  
 %\IRule%  %\IRule% for (;;) { 
 %\IRule%  %\IRule%  %\IRule% dx = Grant.load() - tx ; 
 %\IRule%  %\IRule%  %\IRule% if (dx > 0) return ; 
 %\IRule%  %\IRule%  %\IRule% LocationWait (s) ; 
 %\IRule%  %\IRule% }
 %\IRule% }
 %\IRule%  
 %\IRule% // Optional optimizations ...
 %\IRule% // Try to avoid unnecessary long-term notification operations in the common
 %\IRule% // uncontended case where there are no waiters.
 %\IRule% // Our approach automatically confers the same "fast path" benefits conferred by the
 %\IRule% // well-known "benaphore" optimization :  
 %\IRule% // https://www.haiku-os.org/legacy-docs/benewsletter/Issue1-26.html#Engineering1-26
 %\IRule%  
 %\IRule% void SemaPost () { 
 %\IRule%  %\IRule% auto g = Grant.fetch_add(1) ; 
 %\IRule%  %\IRule%  
 %\IRule%  %\IRule% // Optional optimization ...
 %\IRule%  %\IRule% // In the case where there are no long-term waiting threads on this specific semaphore, 
 %\IRule%  %\IRule% // we would prefer to avoid otherwise unnecessary accesses to the long-term waiting array.   
 %\IRule%  %\IRule% // Absent such an optimization, back-to-back sequences of SemaPost() operations would tend to 
 %\IRule%  %\IRule% // "march" through the long-term waiting array, increasing cache pressure, with futile accesses.  
 %\IRule%  %\IRule% //
 %\IRule%  %\IRule% // To this end, we fetch and examine the value of S%$\rightarrow$%Ticket.  
 %\IRule%  %\IRule% // If the difference between that observed ticket value and the value we just wrote into
 %\IRule%  %\IRule% // grant is less than the LongTermThreshold value then we can safely intuit that there
 %\IRule%  %\IRule% // are no long-term waiters that need notification, and can skip the long-term notification step.
 %\IRule%  %\IRule% // Note that this approach is conservative and racy.
 %\IRule%  %\IRule% // If some other thread calls SemaPost() in the window between the fetch_add(), above, and
 %\IRule%  %\IRule% // the fetch of S%$\rightarrow$%Ticket, then we might still perform unnecessary notification, but
 %\IRule%  %\IRule% // that is benign in terms of correctness. 
 %\IRule%  %\IRule% // That is, the fast-path return does not filter out all unnecessary long-term notifications.
 %\IRule%  %\IRule% // 
 %\IRule%  %\IRule% // This optimization constitutes a "bet".  
 %\IRule%  %\IRule% // The downside is that we incur more coherence traffic as all SemaPost() operations access 
 %\IRule%  %\IRule% // both Ticket and Grant fields.   Having said that, the cache line underlying Ticket and Grant is 
 %\IRule%  %\IRule% // most likely still in local "M"-state (MESI "modified") at the time of the load() and has 
 %\IRule%  %\IRule% // not yet been arbitrated away by concurrent accesses during the fetch_add() - load() window.  
 %\IRule%  %\IRule% // The economic "profitability" model for this optimization is thus platform-specific.  
 %\IRule%  %\IRule% // The access to Ticket is wasted and futile if we still ultimately need to poke
 %\IRule%  %\IRule% // long-duration waiters spinning in the bucket.  
 %\IRule%  %\IRule% // 
 %\IRule%  %\IRule% // If we fail to take the fast-path return, then access to Ticket was futile
 %\IRule%  %\IRule% // and we have incurred a penalty in terms of path length and coherence traffic.
 %\IRule%  %\IRule% // So we have a tension and trade-off.
 %\IRule%  %\IRule% // The fast-path attempt helps under light load but may incur a performance penalty
 %\IRule%  %\IRule% // under contention.  
 %\IRule%  %\IRule% //
 %\IRule%  %\IRule% // Using the fast-path shortcut return optimization means that the
 %\IRule%  %\IRule% // SemaPost() operator must access both Ticket and Grant fields.
 %\IRule%  %\IRule% // This precludes or obviates any coherence traffic benefit we might see from 
 %\IRule%  %\IRule% // sequestering Ticket and Grant fields on distinct cache lines. 
 %\IRule%  %\IRule% int64_t dx = g - Ticket.load() ;        // surplus - excess
 %\IRule%  %\IRule% if (dx >= 0) {
 %\IRule%  %\IRule%  %\IRule% return ;        // common fast-path return : short-cut
 %\IRule%  %\IRule% }
 %\IRule%  %\IRule%  
 %\IRule%  %\IRule% // Poke successor-of-successor from long-term mode into short-term mode
 %\IRule%  %\IRule% // The immediate successor has already been enabled to enter the critical section.
 %\IRule%  %\IRule% // Our approach yields some useful pipeline parallelism as we now act
 %\IRule%  %\IRule% // to shift the successor's successor from long-term waiting to short-term waiting mode. 
 %\IRule%  %\IRule% const auto s = IndexToBucket (TWAHash(g)) ; 
 %\IRule%  %\IRule% LocationWakeAll (s) ; 
 %\IRule% }
 %\IRule%  
 %\IRule% void SemaPostConservative () { 
 %\IRule%  %\IRule% auto g = Grant.fetch_add(1) ; 
 %\IRule%  %\IRule% g += LongTermThreshold ; 
 %\IRule%  %\IRule% const auto s = IndexToBucket (TWAHash(g)) ; 
 %\IRule%  %\IRule% LocationWakeAll (s) ; 
 %\IRule% }
} ;

\end{lstlisting} 

\end{adjustwidth} 
\twocolumn

\end{document}